\documentclass[aip, jap, a4paper, reprint]{revtex4-1}

\usepackage[utf8]{inputenc}

\usepackage{subfig}

\usepackage{amsmath}

\usepackage{boxedminipage}

\usepackage{listings}

\usepackage{ifpdf}
\ifpdf
\usepackage[pdftex]{graphicx}
\else
\usepackage{graphicx}
\fi

\begin{document}
\title{A flexible Bloch mode method for computing complex band structures and impedances of two-dimensional photonic crystals}
\author{Felix J. Lawrence}
\email[]{felix@physics.usyd.edu.au}
\affiliation{CUDOS and Institute of Photonics and Optical Science (IPOS), School of Physics, University of Sydney, NSW 2006, AUSTRALIA}

\author{Lindsay C. Botten}
\author{Kokou B. Dossou}
\affiliation{CUDOS and Department of Mathematical Sciences, University of Technology, Sydney, NSW 2007, AUSTRALIA}

\author{R. C. McPhedran}
\author{C. Martijn de Sterke}
\affiliation{CUDOS and Institute of Photonics and Optical Science (IPOS), School of Physics, University of Sydney, NSW 2006, AUSTRALIA}

\date{\today}

\ifpdf
\DeclareGraphicsExtensions{.pdf, .jpg, .tif}
\else
\DeclareGraphicsExtensions{.eps, .jpg}
\fi

\begin{abstract}
	We present a flexible method that can calculate Bloch modes, complex band structures, and impedances of two-dimensional photonic crystals from scattering data produced by widely available numerical tools.  The method generalizes previous work which relied on specialized multipole and FEM techniques underpinning transfer matrix methods.  We describe the numerical technique for mode extraction, and apply it to calculate a complex band structure and to design two photonic crystal antireflection coatings.  We do this for frequencies at which other methods fail, but which nevertheless are of significant practical interest.
\end{abstract}
\maketitle

\section{Introduction} 
\label{sec:introduction}
When modeling photonic crystals (PCs), it is important to consider all the relevant Bloch modes.  Light at a fixed frequency, polarization, and incident angle exists in a PC as a superposition of a set of propagating and evanescent Bloch modes, the PC's eigenstates.  At low frequencies, only one mode generally needs to be considered.  For light at frequencies above the first Wood anomaly \cite{wood}, each row of holes in the PC diffracts light into several propagating orders, so the PC may support multiple propagating Bloch modes. At the PC's front and back interfaces, some of its modes couple via reflection, affecting the overall reflection and transmission through the PC, so it is important to model all relevant modes.

It is often important to include evanescent modes \cite{Smaali2003443}.  If the PC is not long---for example, if it is a layer in a thin antireflection coating---then evanescent modes can play a role in energy transport \cite{Stefanou:92}.  Evanescent modes can also play a role in field matching across an interface between PCs \cite{Lawrence:2008p79} or PC waveguides \cite{deSterke:09}.  The propagative qualities of an evanescent mode are well-represented by its complex band structure \cite{heine1964}, which augments the traditional band structure, conveying information about the rate at which the mode accumulates phase together with information about the mode's decay rate.

There have been a number of studies seeking to derive impedance-like quantities to characterize reflection at PC interfaces by a scalar \cite{Biswas:2004p465, Smigaj:2011p1695}.  Furthermore, a number of studies have adapted metamaterial parameter extraction techniques \cite{Simovski:2007p1704} to photonic crystals, and used them to design antireflection coatings \cite{Miri:2010p791, Kim:2009p1688}.  However, since these techniques characterize reflection and transmission by a single complex number each, they cannot handle problems involving multiple modes, where every mode reflects into every other mode.  Scalar-based methods generally give manifestly incorrect results for light at frequencies above the first Wood anomaly, which ranges from $a_x/\lambda =1/n$ for normally incident light to $a_x/\lambda = 1/2n$ for light at the Brillouin-zone edge, where $a_x$ is the length of the lattice vector parallel to the interface, $\lambda$ is the free space wavelength and $n$ is the PC's background index.  Above this frequency, generally several Bloch modes must be simultaneously considered in each PC, regardless of whether these modes are propagating or evanescent.  Reflection at a PC/PC interface is well-described by a matrix that maps incident modes to reflected modes, as we have shown previously \cite{Lawrence:2008p79, Lawrence:2009p11}.  In our experience, the minimum acceptable dimension of this reflection matrix, as argued in Sec.~\ref{sub:background_theory}, is usually
\begin{equation}
	\label{eq:numprop}
	M_{\text{min}} = 
	\left\lfloor\frac{a_x}{n\lambda}(1+\sin\theta_i)\right\rfloor + 
	\left\lfloor\frac{a_x}{n\lambda}(1-\sin\theta_i)\right\rfloor + 1,
\end{equation}
where $\theta_i$ is the incident angle from a uniform dielectric with the PC's background index, and $\lfloor x \rfloor$ denotes the \emph{floor} of $x$.
We have previously achieved accurate results modeling PC stacks using impedance matrices of this dimension and higher \cite{Lawrence:2008p79, Lawrence:2009p11, Lawrence:2010p1345}.

A number of methods for finding multiple Bloch modes and complex band structures have been demonstrated.  Transfer-matrix \cite{Gralak:00} and scattering-matrix \cite{Botten:2001p9} based methods were developed to derive a PC's Bloch modes from the properties of a single grating layer.  The plane wave expansion method has also been extended to include evanescent modes \cite{Hsue:2004}.  Finally, Ha \emph{et al.} presented a method for extracting Bloch modes from the output of an EM solver \cite{Ha:2009p1388}, or even near-field measurements \cite{Sukhorukov:09, Ha:2011p2082}.  We improve the accuracy, stability and efficiency of Ha \emph{et al.}'s method and extend it to calculate PC impedances for two-dimensional (2D) PCs, which can be used to calculate reflection and transmission at interfaces \cite{Lawrence:2008p79, Lawrence:2009p11}.  These PC impedances and the reflection and transmission operators are represented by matrices; our method supports the presence and interaction of multiple Bloch modes and so it can work well both above and below the first Wood anomaly.

We have made software available that uses the method described in this paper to calculate PCs' Bloch modes, complex band structures, and impedances.  The software, called BlochCode, can then use these complex band structures and impedances to calculate reflection and transmission matrices and coefficients for arbitrary stacks of PCs. BlochCode is open-source and is available on the internet \cite{blochcodeurl}.

In Sec.~\ref{sec:theory}, we present our method for finding Bloch modes from the electric field $E$ and the magnetic field $H$ in a PC structure.  Sec.~\ref{sub:background_theory} recaps some useful results from our previous work \cite{Lawrence:2009p11} and provides some background theory.  Sec.~\ref{sub:finding_modes} details our improvements to Ha \emph{et al.}'s method \cite{Ha:2009p1388} of finding Bloch factors and modal fields, and Sec.~\ref{sec:numerical_procedure} outlines our procedure for successfully applying this method to minimize the residual derived in Sec.~\ref{sub:finding_modes}. Sec.~\ref{sub:calculating_impedance} explains how we calculate PC impedance matrices from the modal fields.  In Sec.~\ref{sec:application} we apply our method to demonstrate its utility.  In Sec.~\ref{sub:complex_band_structure} we calculate the complex band structure for light normally incident on a triangular lattice PC.  In Sec.~\ref{sub:oldcoating} we reproduce the design process of a known antireflection coating for a PC, at a frequency and incident angle for which it is critical to include at least two Bloch modes in the calculations. Finally, in Sec.~\ref{sub:park_coating} we use our method to design an all-polarization antireflection coating for a square lattice self-collimating PC, at a high frequency where a scalar method cannot find a coating for the PC \cite{Park:2010p651}.

\section{Theory} 
\label{sec:theory}
Our method uses a two-step process to extract a PC's modes and impedance from the field in a finite length of the PC.  The PC is assumed to be two-dimensional, lossless, and to have relative permeability $\mu_r = 1$.  Like Ha \emph{et al.}'s method \cite{Ha:2009p1388}, we could use data generated by FEM or FDTD simulations, or even experimentally measured by a near-field probe such as a SNOM \cite{Ha:2011p2082}, although the impedance part of our method is not valid for SNOM data, which is derived from a 3D object.  First, the Bloch factors and the Bloch modal fields are found (Sec.~\ref{sub:finding_modes}), then these modes are analyzed to calculate the PC's impedance (Sec.~\ref{sub:calculating_impedance}).

\subsection{Background Theory} 
\label{sub:background_theory}
Two-dimensional PCs in the $x-y$ plane may be described as a stack of gratings parallel to the $x$ axis \cite{Botten:2000p505}, each of which diffracts incident light into an infinite set of grating orders.  At the edge of each unit cell, the PC's Bloch modes may be written as a superposition of the underlying grating orders \cite{Botten:2001p9}.  Their directions are given by the grating equation
\begin{equation}
	\label{eq:grating}
	k_x^{(p)} = k_x + \frac{2 \pi p}{a_x} = k \sin\theta_i + \frac{2 \pi p}{a_x},
\end{equation}
where $k_x$ is the $x$ component of the incident plane wave's wavevector, $k_x^{(p)}$ is that of the $p$th diffraction order, and $a_x$ is the length of the lattice vector parallel to the $x$-axis.  The wavevector component in the direction perpendicular to the grating is $k_y^{(p)} = \sqrt{k^2 - {k_x^{(p)}}^2}$ where $k$ is the wavenumber in the medium.  Evanescent grating orders have imaginary $k_y^{(p)}$, so for a given $k$ and $k_x^{(p)}$, the number of propagating grating orders is the number of solutions to Eq.~\eqref{eq:grating} with real $k_y^{(p)}$, or $M_\text{min}$ in Eq.~\eqref{eq:numprop}.  In our experience, $M_\text{min}$ also provides an upper bound on the number of propagating Bloch modes, and at non-normal incidence is a lower bound on the number of Bloch modes required to model a PC accurately.  At normal incidence, symmetry allows odd modes to be ignored, so in this case good results may be obtained with fewer than $M_{\text{min}}$ modes---see Sec.~\ref{sub:park_coating}.  Using Bloch modes found from accurate multipole and FEM transfer matrix methods \cite{McPhedran:2000p784, Botten:2004p5}, we have consistently had success modeling PCs with no more than $M_\text{min} + 2$ Bloch modes.

Bloch's theorem relates the electric and magnetic fields associated with each mode at equivalent points in different unit cells of a PC.  The ratio of each mode's field at points separated by the lattice vector $\mathbf{e}_1 = (a_x, 0)$ is $e^{i k_x a_x}$.  For the PC's other lattice vector $\mathbf{e}_2$, this ratio is different for each mode and is the mode's Bloch factor, denoted by $\mu$.  Calculating $\mu$ for each mode is the goal of Sec.~\ref{sub:finding_modes}.  For square and rectangular lattices, $\mathbf{e}_2 = (0, a_y)$ and $\mu = e^{ik_y a_y}$, where $k_y$ is the $y$ component of the mode's wavevector.  For triangular lattices, the lattice vector $\mathbf{e}_2$ is $(a_x/2, a_y)$ and so the Bloch factor may be written $\mu = e^{i(k_x a_x/2+k_y a_y)}$.

Bloch modes come in forward/backward pairs.  Popov \emph{et al.} provide a useful discussion of symmetry properties \cite{Popov:1986p1233}.  We assume mirror symmetry in each unit cell, which means that each backward mode's field profile in a unit cell is the reflection on the $x$-axis of its forward partner's.  The Bloch factors of a pair are related because of this: for square and rectangular lattices, $\mu_b = 1/\mu_f$, where $\mu_f$ and $\mu_b$ are respectively the Bloch factors of the forward and backward modes.  For triangular-like lattices, the symmetry is more complicated since the reflection of $\mathbf{e}_2$ is not $-\mathbf{e}_2$, the translation corresponding to the field ratio $1/\mu_f$, but $(a_x/2, -a_y)$; these vectors differ by $-\mathbf{e}_1$. Accounting for this discrepancy, we find $\mu_b = e^{-i k_x a_x} / \mu_f$ for triangular lattices.

A PC's impedance is defined in terms of two matrices, $\mathbf{E}$ and $\mathbf{H}$ \cite{Lawrence:2009p11}.  For $E = E_z$ polarized light, each matrix maps a vector of forward Bloch mode amplitudes $\mathbf{c}_+$ to a vector of the $E_z$ or $H_x$ fields associated with each grating diffraction order.  
Specifically, $E_{p,m}$, the $(p,m)$th element of $\mathbf{E}$, is the $E_z$ field of normalized mode $m$ due to forward and backward plane waves in grating order $p$, at the centre ($x=0$) of a unit cell's edge.
Thus, for a set of forward propagating/decaying Bloch modes $\mathbf{c}_+$, the field components along the edge of the unit cell, i.e., the quantities that are continuous across an interface between PCs or dielectrics, are
\begin{equation}
	\label{eq:eh_demonstration}
	E_z(x) = \sum_p \mathbf{E}_p~\!\mathbf{c}_+ e^{i k_x^{(p)}x},~H_x(x) = \sum_p \mathbf{H}_p~\!\mathbf{c}_+ e^{i k_x^{(p)}x},
\end{equation}
where $\mathbf{E}_p$ and $\mathbf{H}_p$ are the rows of $\mathbf{E}$ and $\mathbf{H}$ corresponding to grating order $p$.  In the $H = H_z$ polarization, $\mathbf{E}$ and $\mathbf{H}$ map to $E_x$ and $H_z$ fields, and these quantities replace $E_z$ and $H_x$ in Eq.~\eqref{eq:eh_demonstration}.

Previously \cite{Lawrence:2009p11}, we defined PC impedances in terms of these matrices.  For $E_z$ polarized light, the impedance of a PC is
\begin{equation}
	\label{eq:Z_Ez}
	{\cal Z} = {\mathbf{H}_0}^T (\mathbf{I} + \mathbf{Q}) \mathbf{E} + {\mathbf{E}_0}^T (\mathbf{I} - \mathbf{Q}) \mathbf{H},
\end{equation}
and for $H_z$ polarized light it is
\begin{equation}
	\label{eq:Z_Hz}
	{\cal Z} = -\left({\mathbf{H}_0}^T (\mathbf{I} - \mathbf{Q}) \mathbf{E} + {\mathbf{E}_0}^T (\mathbf{I} + \mathbf{Q}) \mathbf{H}\right),
\end{equation}
where $\mathbf{E}$ and $\mathbf{H}$ are calculated for the PC, and $\mathbf{E}_0$ and $\mathbf{H}_0$ are calculated for a reference material, usually free space.  $\mathbf{Q}$ is a diagonal matrix that takes into account the half-period shift of gratings in triangular lattice PCs: for square lattices $\mathbf{Q} = \mathbf{I}$, and for triangular lattices $\mathbf{Q} =\text{diag}((-1)^p)$, where $p$ is the grating order.

Given impedances $\mathcal{Z}_1$ and $\mathcal{Z}_2$ for two PCs, it is simple to calculate the reflection and transmission matrices across their interface \cite{Lawrence:2009p11}:
\begin{subequations}
	\label{eq:r_t_of_Z}
  \begin{eqnarray}
    \mathbf{T}_{12} &=& (\mathbf{A}_{12}^T \mathbf{A}_{12} + \mathbf{I})^{-1} 2 \mathbf{A}_{12}^T, \label{eq:t12}\\
    \mathbf{R}_{12} &=& (\mathbf{A}_{12} \mathbf{A}_{12}^T + \mathbf{I})^{-1} (\mathbf{A}_{12} \mathbf{A}_{12}^T - \mathbf{I}), \label{eq:r12}
  \end{eqnarray}
\end{subequations}
where $\mathbf{A}_{12} = \mathcal{Z}_1^{-1} \mathcal{Z}_2$.

\subsection{Finding modes} 
\label{sub:finding_modes}
Our method of finding the Bloch modes and Bloch factors is based on the method presented by Ha \emph{et al.} \cite{Ha:2009p1388}, although our method offers some significant improvements in accuracy and efficiency.  We take field data for several unit cells of a PC, and try to write it as a superposition of Bloch modes, thus finding the modal fields and Bloch factors.  The final steps of our mode-finding method impose symmetry relationships between forward and backward modal fields, increasing accuracy by almost halving the number of unknowns in the problem.  We now outline our method.

\begin{figure}[htbp]
	\centering
		\includegraphics[]{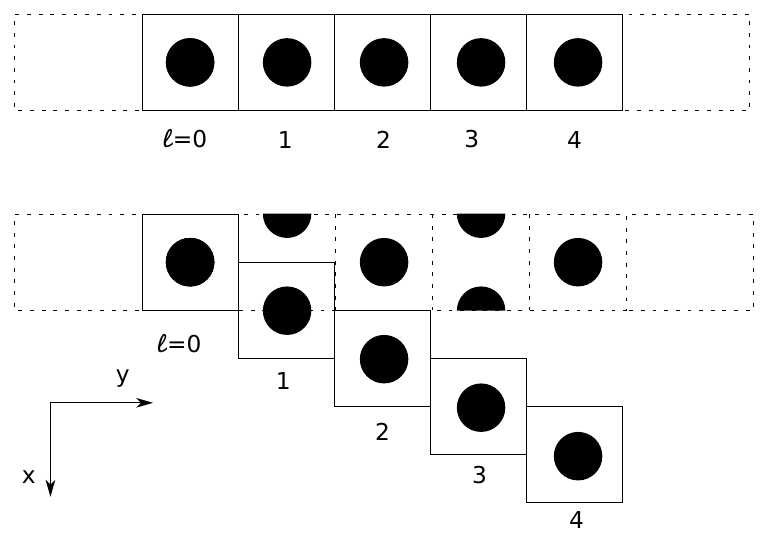}
	\caption{Schematic of $L=5$ PC structures for a square and a triangular PC lattice.  The squares with solid edges are the unit cells used by our method.  For the triangular lattice PC, the field in the solid-edge unit cells are calculated from the unit cells of the simulated structure (dashed edges) using Bloch's theorem, with the ratio $e^{i k_x a_x}$ between adjacent cells' fields.}
	\label{fig:simmoschem}
\end{figure}

In an EM solver, we simulate a section of 2D PC with Bloch-Floquet periodic boundary conditions on two boundaries, and uniform dielectric on the others (Fig.~\ref{fig:simmoschem}).  We sample the $E_z$ or $E_x$ (depending on polarization) field component at many ($N_p$) points in unit cell $\ell = 0$, and then at the equivalent points in each of the other unit cells.  If desired, $E_y$, $H_x$, $H_y$, or $H_z$ may be used in place of or in addition to $E_z$ and $E_x$.  For triangular lattice PCs, we use the field in the simulated unit cells (dashed edges in Fig.~\ref{fig:simmoschem}) to calculate the field in the unit cells separated by a lattice vector (solid edges); we apply Bloch's theorem with integer multiples of the lattice vector $(a_x, 0)$.

We seek to write these electric field components as a superposition of forward and backward Bloch modes.   So we want to express every $U_\ell(\mathbf{r})$, i.e., the $E_z$ or $E_x$ field component for sampled point $\mathbf{r}$ in unit cell $\ell$, as
\begin{equation}
	U_\ell(\mathbf{r}) = \sum_{m} \mu_m^\ell A_m(\mathbf{r}) + \sum_{m^\prime} (1/{\mu_{m^\prime}}^{L-1-\ell}) A_{m^\prime}(\mathbf{r})  + w(\ell, \mathbf{r}),
	\label{eq:EAMu}
\end{equation}
where  $A_m(\mathbf{r})$ and $\mu_m$ are respectively the modal field and the Bloch factor of forward mode $m$; $m^\prime$ denotes backward modes, and $w(\ell,\mathbf{r})$ is the residual error.  More specifically, for forward modes, $A_m(\mathbf{r})$ is the field component of mode $m$ at point $\mathbf{r}$ of the first unit cell, $\ell = 0$.  The Bloch factor $\mu_m$ is the ratio of the field in cells $\ell + 1$ and $\ell$, so $\mu_m^\ell A_m(\mathbf{r})$ is the field component of forward mode $m$ at point $\mathbf{r}$ of unit cell $\ell$.  To avoid ill-conditioning, the field $A_{m^\prime} (\mathbf{r})$ at point $\mathbf{r}$ of each backward mode $m^\prime$ is defined in the last unit cell, $\ell = L-1$.  This means that the coefficients of $A_m(\mathbf{r})$ and $A_{m^\prime}(\mathbf{r})$ in Eq.~\eqref{eq:EAMu} have moduli no greater than 1.  As noted in Sec.~\ref{sub:background_theory}, the Bloch factor $\mu_{m^\prime}$ of each backward mode is related to that of its forward partner; we enforce this relationship in practice, thereby halving the number of Bloch factors that must be found.

Equation \eqref{eq:EAMu} for all $\ell$ and all sampled $\mathbf{r}$ may be written in matrix form as:
\begin{equation}
	\mathbf{U} = \mathbf{C} \mathbf{A} + \mathbf{W},
	\label{eq:CA_eq_U}
\end{equation}
where $\mathbf{U}$ contains the $E_z$ or $E_x$ field components from the EM solver, $\mathbf{A}$ is a matrix of modal fields, $\mathbf{C}$ is a matrix constructed from Bloch factors, and $\mathbf{W}$ is a matrix of residuals $w(\ell,\mathbf{r})$ that must be minimized.  $\mathbf{U}$ is a $L \times N_p$ matrix: the field in its $\ell$th row and $r$th column is $U_{\ell, r}= U_\ell(\mathbf{r})$, the field component at point $\mathbf{r}$ in unit cell $\ell$.  Similarly, $\mathbf{A}$ is a $M \times N_p$ matrix; the field in its $m$th row and $r$th column is $A_{m,r} = A_m(\mathbf{r})$, the field of mode $m$ at point $\mathbf{r}$ in cell $\ell = 0$ for forward modes, or cell $\ell = L-1$ for backward modes.  $\mathbf{C}$ is a $L \times M$ matrix. For a forward mode $m$, the $(\ell,m)$th element of $\mathbf{C}$ is ${\mu_m}^\ell$, and for a backward mode $m^\prime$, the $(\ell,m^\prime)$th element is ${1/\mu_{m^\prime}}^{L-1-\ell}$.  If multiple field components (e.g. $E_z$, $H_x$ and $H_y$) are to be used to find the modes, then the additional data can be added as extra columns in $\mathbf{U}$.

We start the optimization process knowing $\mathbf{U}$, and with information about the structure of $\mathbf{C}$, and no direct information about $\mathbf{A}$.  In our method, we first find the Bloch factors that determine $\mathbf{C}$, a relatively difficult problem. Once $\mathbf{C}$ is known, solving Eq.~\eqref{eq:CA_eq_U} for the modal fields $\mathbf{A}$ becomes a pure least-squares problem that can be solved accurately and efficiently using standard techniques.

To find the modes, we seek to minimize the difference between the observed field $\mathbf{U}$ and the superposition of Bloch mode fields $\mathbf{CA}$.  That is, we seek to minimize $||\mathbf{W}||_F^2$ in Eq.~\eqref{eq:CA_eq_U}, the sum of squared moduli of the elements of $\mathbf{W}$.  Constraining the problem by dividing by the squared Frobenius norm $||\mathbf{U}||_F^2$ of $\mathbf{U}$, the quantity we minimize is
\begin{equation}
	\label{eq:CA_residual}
	w^2 = \frac{||\mathbf{U} - \mathbf{CA}||_F^2}{||\mathbf{U}||_F^2},
\end{equation}
where $w^2 = ||\mathbf{W}||_F^2/||\mathbf{U}||_F^2$.
First we eliminate $\mathbf{A}$ from Eq.~\eqref{eq:CA_residual} in order to find $\mathbf{C}$ with a numerical minimizer.
We use an alternative representation of the Frobenius norm, $||\mathbf{U}||_F = \sqrt{\text{tr}(\mathbf{U}^H \mathbf{U})}$, to write
\begin{equation}
	\label{eq:CA_residual_factorised}
	w^2 = \frac{{\text{tr}((\mathbf{U}^H - \mathbf{A}^H \mathbf{C}^H)(\mathbf{U} - \mathbf{CA}))}}
		{||\mathbf{U}||_F^2}.
\end{equation}
Finding $\mathbf{A}$ for arbitrary $\mathbf{C}$ is a standard least-squares problem; the optimal $\mathbf{A}$ satisfies $\mathbf{C}^H \mathbf{CA}=\mathbf{C}^H \mathbf{U}$.  We expand Eq.~\eqref{eq:CA_residual_factorised}, twice apply this relation, and rearrange to get
\begin{equation}
	\label{eq:C_residual}
	w^2 = 1 - \frac{\text{tr}(\mathbf{U}^H \mathbf{CC}^+\mathbf{U})}{||\mathbf{U}||_F^2},
\end{equation}
where $\mathbf{C}^+ = (\mathbf{C}^H \mathbf{C})^{-1} \mathbf{C}^H$ is the Moore-Penrose pseudoinverse of $\mathbf{C}$.

Using Eq.~\eqref{eq:C_residual} and a numerical minimizer, the Bloch factors that determine $\mathbf{C}$ may often be found to a useful level of accuracy (see Sec.~\ref{sec:numerical_procedure} for implementation details).  In order to improve the accuracy and reliability of the results, we impose further physical constraints.

The PC impedance method \cite{Lawrence:2008p79, Lawrence:2009p11} assumes the unit cell to be up-down symmetric, which causes the forward and backward modes to be related.  So far, we have only imposed a relationship between the forward and backward Bloch factors, not the modal fields within each unit cell.  We can halve the number of unknowns in $\mathbf{A}$ and strongly improve the quality of our results by enforcing this relationship in the minimization process.

We commence by partitioning the forward ($f$) and backward ($b$) modes, and the points in the left ($L$; $y \leq a_y/2$) and right ($R$; $y \geq a_y/2$) halves of the unit cell: 
\begin{subequations}
	\begin{equation}
		\mathbf{U} = \left(\mathbf{U}_L, \mathbf{U}_R\right),~\mathbf{C} = \left(\mathbf{C}_f, \mathbf{C}_b\right),
	\end{equation}
	\begin{equation}
		\mathbf{A} =
		\begin{pmatrix}
			\mathbf{A}_{L,f} & \mathbf{A}_{R, f}\\
			\mathbf{A}_{L,b} & \mathbf{A}_{R, b}
		\end{pmatrix}.
	\end{equation}
\end{subequations}
After normalization, the field of a backward mode is the field of its forward partner reflected about the $x$-axis, thus 
\begin{equation}
	\label{eq:A_constraint}
	\left(\mathbf{A}_{L,b}, \mathbf{A}_{R, b}\right) = 
	\left(\gamma \mathbf{A}_{R,f}\mathbf{P}, \gamma \mathbf{A}_{L,f} \mathbf{P}^{-1}\right),
\end{equation}
where $\mathbf{P}$ is the permutation matrix that maps points $(x,a_y - y)$ to $(x, y)$, and $\gamma$ is a normalizing diagonal matrix whose elements are the ratio of backward and forward mode amplitudes.  
The columns of $\mathbf{A}_{R,f}$ and $\mathbf{A}_{R,b}$, corresponding to points in the right half of the unit cell, can easily be ordered so that $\mathbf{P} = \mathbf{I}$; from now on we assume this ordering.
Eq.~\eqref{eq:CA_eq_U} can now be written with roughly half as many unknowns,
\begin{equation}
	\label{eq:CpCm_multiplied}
	\left(\mathbf{U}_L, \mathbf{U}_R \right) =
	\left(\mathbf{C}_f, \mathbf{C}_b \gamma \right)
	\begin{pmatrix}
		\mathbf{A}_{L,f} & \mathbf{A}_{R, f}\\
		 \mathbf{A}_{R,f} & \mathbf{A}_{L, f}
	\end{pmatrix}
	+ \mathbf{W}.
\end{equation}
$\mathbf{C}_b \gamma$ represents each backward mode's amplitude in each cell, relative to that of the corresponding forward mode in cell 0.

The constraints on $\mathbf{A}$ (Eq.~\eqref{eq:A_constraint}) mean that Eq.~\eqref{eq:CpCm_multiplied} does not have a least-squares form, so may not be immediately simplified in the way that Eq.~\eqref{eq:CA_residual} led to Eq.~\eqref{eq:C_residual}.  To transform Eq.~\eqref{eq:CpCm_multiplied} into a more useful form, we block-diagonalize $\mathbf{A}$ and right-multiply by the matrix $
\left(\begin{smallmatrix}
	\mathbf{I} & \mathbf{I} \\
	\mathbf{I} & -\mathbf{I}
\end{smallmatrix}\right)$,
 to show
\begin{equation}
	\label{eq:UpUm}
	(\mathbf{U}_+,~\mathbf{U}_-) = (\mathbf{C}_+ \mathbf{A}_+,~\mathbf{C}_- \mathbf{A}_-)	+ \mathbf{W^\prime}.
\end{equation}
Here we have introduced the symmetric and antisymmetric forms $\mathbf{U}_\pm = \mathbf{U}_L \pm \mathbf{U}_R$, $\mathbf{C}_\pm = \mathbf{C}_f \pm \mathbf{C}_b \gamma$, and $\mathbf{A}_\pm = \mathbf{A}_{L,f} \pm \mathbf{A}_{R,f}$.

Eq.~\eqref{eq:UpUm} takes the form of two independent least-squares equations, each with half the dimension of Eq.~\eqref{eq:CpCm_multiplied}.  The two equations must be satisfied simultaneously, so to find the Bloch factors we can minimize
\begin{equation}
	w^2 = \frac{||\mathbf{U}_+ - \mathbf{C}_+ \mathbf{A}_+||_F^2 + ||\mathbf{U}_- - \mathbf{C}_- \mathbf{A}_-||_F^2}
	{||\mathbf{U}_+||^2_F + ||\mathbf{U}_-||^2_F},
\end{equation}
or equivalently
\begin{equation}
	\label{eq:final_minimise}
	w^2 = 1 - \frac{\text{tr}(\mathbf{U}_+^H \mathbf{C}_+ \mathbf{C}_+^+ \mathbf{U}_+) +
									\text{tr}(\mathbf{U}_-^H \mathbf{C}_- \mathbf{C}_-^+ \mathbf{U}_-)}
						{||\mathbf{U}_+||_F^2 + ||\mathbf{U}_-||_F^2}.
\end{equation}
Again, this quantity may be minimized by a numerical optimizer.  The residual $w^2$ for any solution to Eq.~\eqref{eq:final_minimise} is equal to the residual obtained by inserting the solution into Eq.~\eqref{eq:C_residual}: the two equations differ only in the symmetry constraint on backward modal fields (Eq.~\eqref{eq:A_constraint}). Compared to Eq.~\eqref{eq:C_residual}, we have removed $N_p M$ unknowns from $\mathbf{A}$ (where $N_p \gg M$ is the number of sampled points in each unit cell), halving its dimension at the cost of adding $M$ unknowns to $\mathbf{C}_\pm$ as $\gamma$.  These new unknowns must be found simultaneously with the Bloch factors using a numerical minimizer, so it is important to supply a good starting estimate; our method for doing so is detailed in Sec.~\ref{sec:numerical_procedure}.

\subsection{Calculating impedance} 
\label{sub:calculating_impedance}
Once the Bloch factors and $\gamma$ are known, the modal fields can be reconstructed and analyzed to determine the PC's impedance.  The essential quantities for this calculation are the $E$ and $H$ field components in the plane of the PC interface (i.e., $E_z$ and $H_x$, or $E_x$ and $H_z$, depending on polarization) of each Bloch mode $m$ along the left edge ($y=0$) of a unit cell (see Fig.~\ref{fig:simmoschem}).  These quantities, $E_m(x)$ and $H_m(x)$, may be found from Eq.~\eqref{eq:UpUm} using the known values for $\mathbf{C}_+$ and $\mathbf{C}_-$ and inserting the appropriate $E$ or $H$ fields into $\mathbf{U}_+$ and $\mathbf{U}_-$.

To calculate the impedance, we find the $\mathbf{E}$ and $\mathbf{H}$ matrices for the PC, as defined in Sec.~\ref{sub:background_theory}.  Inserting multiples of unit vectors $\mathbf{c}_+$ into Eq.~\eqref{eq:eh_demonstration}, we can show that
\begin{subequations}
	\begin{equation}
		E_m(x) = {\cal A}_m \sum_{p} E_{p,m}~e^{i k_x^{(p)} x},
	\end{equation}
	\begin{equation}
		H_m(x) = {\cal A}_m \sum_{p} H_{p,m}~e^{i k_x^{(p)} x},
	\end{equation}
\end{subequations}
where ${\cal A}_m$ is the amplitude of the normalized mode $m$, and $E_{p,m}$ and $H_{p,m}$ are the elements of $\mathbf{E}$ and $\mathbf{H}$.
It is straightforward to exploit the orthogonality of the plane wave grating diffraction orders to show that

\begin{subequations}
	\label{eq:EH_elements}
	\begin{equation}
		{\cal A}_m E_{p,m} = 1/a_x \int_{-a_x/2}^{a_x/2} \! E_m(x) e^{-i k_x^{(p)} x} \, dx,\\
	\end{equation}
	\begin{equation}
		{\cal A}_m H_{p,m} = 1/a_x \int_{-a_x/2}^{a_x/2} \! H_m(x) e^{-i k_x^{(p)} x} \, dx.
	\end{equation}
\end{subequations}
Eqs.~\eqref{eq:EH_elements} let us calculate each element of the $\mathbf{E}$ and $\mathbf{H}$ matrices, up to a normalization constant ${\cal A}_m$ per column.  We remove the constants by calculating the PC's impedance (Eq. \eqref{eq:Z_Ez} or \eqref{eq:Z_Hz}) with the PC itself as the reference material: by reciprocity-derived Bloch mode orthogonality relations \cite{Lawrence:2009p11}, this quantity should be the identity matrix.  The diagonal entries of this matrix are the ${{\cal A}_m}^2$; the off-diagonal terms, which should be zero, provide an error estimate.  After normalizing the $\mathbf{E}$ and $\mathbf{H}$ matrices for the PC, we calculate its impedance matrix $\cal Z$ from Eq. \eqref{eq:Z_Ez} or \eqref{eq:Z_Hz} using a reference medium such as free space.
\section{Numerical Procedure} 
\label{sec:numerical_procedure}
Having outlined the theoretical basis of our method for finding the Bloch factors and impedance of a PC at a given frequency, incident angle, and polarization, we now provide some practical detail about our implementation of the method.  We outline the procedure for $M = 3$ pairs of Bloch modes.

In COMSOL Multiphysics 4.2, we simulate a $1 \times 8$ unit cell sample of PC, embedded in its background dielectric, with Bloch-Floquet periodic boundary conditions along the two long boundaries (Fig.~\ref{fig:simmoschem} shows a $1 \times 5$ structure).  Eq. \eqref{eq:UpUm} is a set of $LN_p$ equations, with $2M$ and $M N_p$ unknowns in $\mathbf{C}_\pm$ and $\mathbf{A}_\pm$ respectively.  To be overspecified, the method requires $LN_p > M N_p + 2M$; thus $L=8$ periods and a large $N_p$ is sufficient to find $M=3$ modes. A deeper structure with more unit cells does not necessarily provide useful information about additional evanescent modes, as their amplitude deep inside the structure may be negligible.  From COMSOL we export the relevant $E$ and $H$ field components in the $L=8$ unit cells, sampled over a $101 \times (50L+1)$ grid.

In order to compute a mode, it must be present in the structure with sufficient amplitude to be detected.  Light at normal incidence often fails to excite odd Bloch modes; these \emph{uncoupled modes} \cite{Sakoda:1995p1955} consequently cannot be found by an optimization, which loses accuracy in searching for modes that are not present.  At frequencies above the first Wood anomaly, the frequencies at which the higher order modes are most important, this problem may be avoided by exciting the PC slab not with a normally incident plane wave, but with the first grating diffraction order.  This technique is used in Sec.~\ref{sub:complex_band_structure} and Sec.~\ref{sub:park_coating}.  If the uncoupled mode is not relevant to a particular problem, it may instead be ignored.

If we seek to find $M=3$ Bloch modes, then finding a global minimum of Eq.~\eqref{eq:final_minimise} involves searching for $2M = 6$ complex numbers.  This is a hard problem if attacked directly, but we use an algorithm that gives more consistent success by providing a good starting estimate.  We start by minimizing the residual $w^2$ in Eq.~\eqref{eq:C_residual}, which forces a relationship between forward and backward Bloch factors but not the modal fields.  This involves finding only $M$ complex numbers.  As a starting estimate for the forward Bloch factors, we either take the result of a neighboring simulation, or the analytically calculated Bloch factors for the dielectric background of the PC.  At every step of the minimization, evanescent modes are sorted into forward and backward decaying modes, based on the moduli of their Bloch factors.  The minimization can be done by any standard numerical minimizer, such as SciPy's \cite{scipy} \verb`fmin`, which is a modified Nelder-Mead optimization \cite{Wright:1996:DSM}. At this point, the results are equivalent to those from the method of Ha \emph{et al.} \cite{Ha:2009p1388}, except that we have lessened the likelihood of $\mathbf{C}$ being ill-conditioned by renormalizing the backward Bloch factors $\mu_{m^\prime}$ in Eq. \eqref{eq:EAMu} and setting their phase origin to the end of the PC.

Occasionally, we encounter an instability in which a pair of modes have very large equal and opposite field amplitudes and very small Bloch factors.  When this occurs, we follow a Gram-Schmidt-like process: we subtract the field of non-problematic modes (i.e., modes with $|\mu| > 10^{-3}$) from $\mathbf U$ and repeatedly minimize Eq.~\eqref{eq:C_residual} to find each of the remaining modes individually.

Using the solution to Eq.~\eqref{eq:C_residual} as our estimate for the Bloch factors, the modal fields may be found with a least-squares optimization.  The average field ratio of each pair of backward and forward modes gives us an estimate for $\gamma$.  We now have a plausible estimate for $\gamma$ and the Bloch factors, which we can use as a starting estimate to minimize Eq.~\eqref{eq:final_minimise}.

To further refine the estimates, we repeatedly iterate through the modes, fixing all but one $\mu$ and the corresponding element of $\gamma$, minimizing Eq.~\eqref{eq:final_minimise} to find the two variables.  After this process, we finally minimize Eq.~\eqref{eq:final_minimise} across all 6 complex dimensions simultaneously to obtain the correct Bloch factors and modal fields from which we calculate impedances.  Forward and backward propagating modes are sorted based on their flux \cite{Botten:2001p9}, before impedances are calculated as outlined in Sec.~\ref{sub:calculating_impedance}.

\section{Applications} 
\label{sec:application}
We now apply our method to a range of typical problems.  Each of these problems involves frequencies above the first or second Wood anomaly---frequencies at which scalar methods fail and multiple modes are required to describe the system.  BlochCode, software that implements our method in Python, using SciPy \cite{scipy} and Sage \cite{sage}, is freely available on the internet \cite{blochcodeurl}; we use it here.
\subsection{Complex band structure} 
\label{sub:complex_band_structure}
The first application of our method is to calculate the complex band structure of a PC.  The PC is a triangular lattice of circular air holes with radius $r = 0.3~a$ and lattice constant $a_x = a$ in a dielectric background with $n=3$.  We calculate the band structure for light polarized with the $\mathbf{H}$ field out of the PC plane ($H_z$ polarization) at frequencies $a/\lambda \in (0,0.5)$ in the $\Gamma-M$ direction, i.e., at normal incidence.  Using COMSOL, we calculate the field in an 8 period slab of the PC, and we apply our method to find the largest three Bloch factors.  $w^2$ varies: it is less than $10^{-8}$ at low frequencies and less than $10^{-4}$ at high frequencies.

\begin{figure}[htbp]
	\centering
	\includegraphics[]{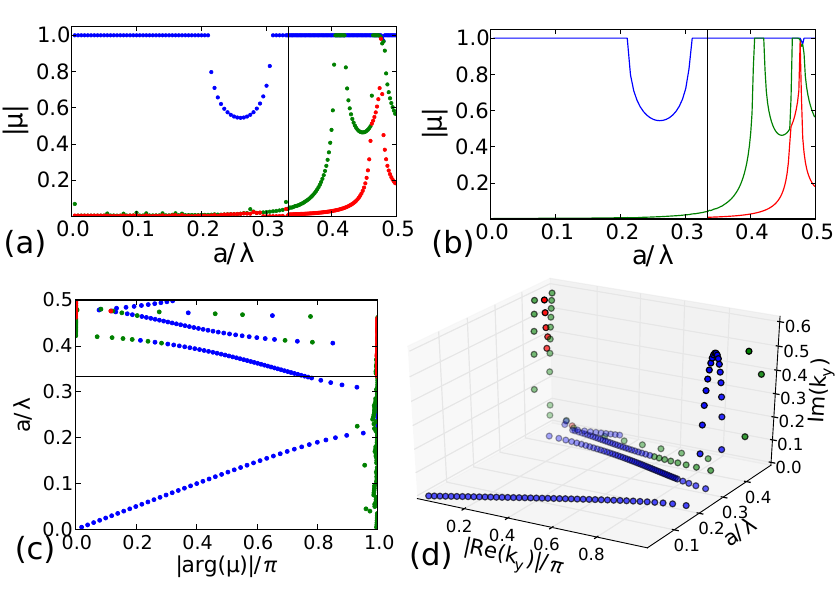}
	\caption{(Color online) Complex band structure for the PC.  The Wood anomaly ($a/\lambda = 0.333$) is marked.  The modes are sorted into colors by $|\mu|$; where two modes are propagating (i.e., have $|\mu| = 1$), they are sorted by $|\text{arg} (\mu)|$.  (a) Magnitude of Bloch factors $|\mu|$, with three Bloch modes found at all frequencies. (b) $|\mu|$ with two Bloch modes found below the Wood anomaly, three above. (c) Argument of Bloch factors. (d) Complex band structure in 3D.}
	\label{fig:bandstructure}
\end{figure}

Fig.~\ref{fig:bandstructure} summarizes the propagation properties of the two/three most dominant modes.  The moduli of the Bloch factors $|\mu|$, which quantify how the modes' amplitudes vary with propagation, are shown in Figs.~2(a) and 2(b).  Below the Wood anomaly, an inspection of $\mathbf{A}$ and $\gamma$ shows that the third mode is barely excited by the normally incident plane wave, and this reduces the accuracy of the results (Fig.~2(a)). Ignoring the uncoupled mode at low frequencies (where the $p=1$ grating order is evanescent and so may not be used to excite the structure, as mentioned in Sec.~\ref{sec:numerical_procedure}) increases the accuracy of the other two modes (Fig.~2(b)).  The complex arguments of the Bloch factors, which quantify how phase is acquired through propagation, are shown in Fig.~2(c), and the information about amplitude and phase is summarized in a single plot in Fig.~2(d).  Aside from slight errors in the phase of strongly evanescent modes in Fig.~2(c), there is good agreement between Fig.~\ref{fig:bandstructure} and Bloch factors calculated by highly accurate multipole techniques.

Figure~\ref{fig:bandstructure} shows that at frequencies below the Wood anomaly there is at most one propagating Bloch mode, which becomes evanescent in the first bandgap with a decay factor $|\mu|$ of no less than 0.5; it still decays far more slowly than the other evanescent Bloch modes at that frequency.  Fig.~2(c) shows that for the evanescent modes, either 0 or $\pi$ phase is acquired across each unit cell.
\subsection{Antireflection coating} 
\label{sub:oldcoating}
Our next application is to reproduce the design of an antireflection coating we presented previously \cite{Lawrence:2009p11}, found using PC impedances calculated with a specialized transfer-matrix method \cite{Botten:2004p5}.  As in this previous paper, our design strategy is to try out a very large number of potential coatings, and choose the coating that gives the lowest reflectance off the coated structure.  The use of PC impedances makes this a feasible problem, as the evaluation of each coating is quick, involving a few operations on $M \times M$ (here $3\times 3$) matrices.

The target PC is a triangular lattice with lattice constant $a_x = a$, consisting of air holes in a dielectric background with $n=2.86$.  The holes are cylinders with radius $r = 0.25~a$.  We seek to coat the PC to minimize reflection for light with frequency $a/\lambda = 0.38$, incident from air at an angle of $30^\circ$ in the $E_z$ polarization.  At this frequency and incident angle, $M_\text{min} = 2$; we consider a total of 3 modes to ensure accuracy.  As in our previous work \cite{Lawrence:2009p11}, we seek a two-layer coating, where the degree of freedom is $a_y$, the lattice vector component perpendicular to the air/PC interface.  For a regular triangular lattice, $a_y = \frac{\sqrt{3}}{2} a$.

We choose 121 candidate PCs with $a_y \in [0.6,1.8]~\frac{\sqrt{3}}{2} a$ and simulate 8 periods of each in COMSOL.  We apply our method to the resulting data, using the Bloch factors of the previous PC as the starting estimate for the next.  BlochCode processes the 121 PCs in approximately 13 minutes on a 3.06~GHz Intel Core~2~Duo desktop computer.  An equivalent approach that only requires one PC to be evaluated is detailed in Sec.~\ref{sub:park_coating}; we do not use it here since the purpose of this section is to demonstrate the reliability and consistency of the optimization procedure.

\begin{figure}[htbp]
\centering
	\includegraphics[]{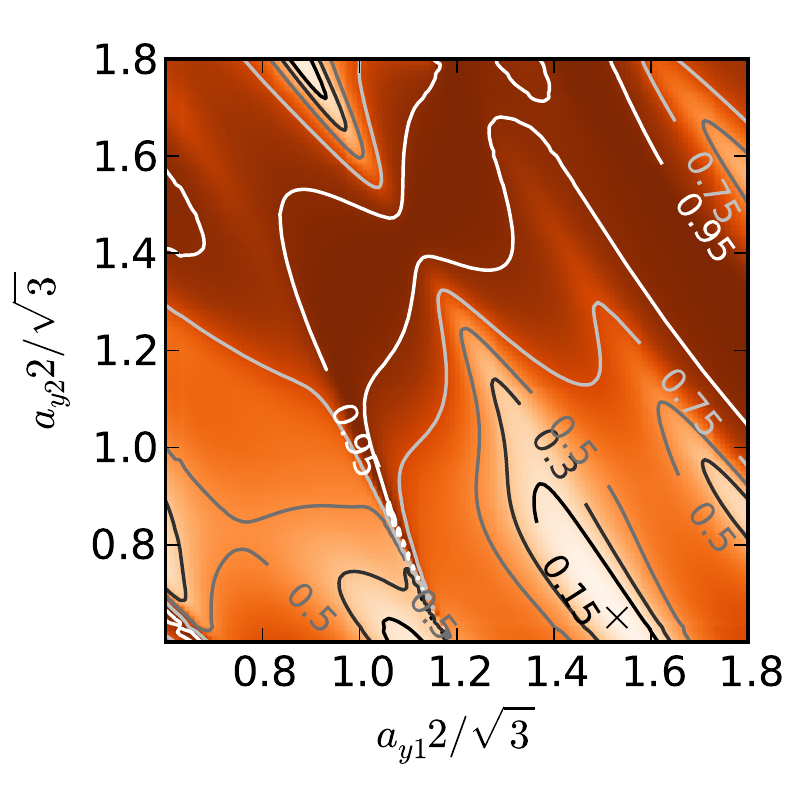}
	\caption{(Color online) Reflectance of the coated PC as a function of $a_{y1}$ and $a_{y2}$, the relative thicknesses of the two coating layers, calculated using PC impedances from BlochCode.  The minimum reflectance is marked.}
\label{fig:pracoat}
\end{figure}

We then calculate the reflectances off the $121^2 = 14641$ coated stacks (Fig.~\ref{fig:pracoat}), which takes 34 seconds on a single core of the desktop computer.  The optimal coating is found to have thicknesses $a_{y1} = 1.53~\frac{\sqrt{3}}{2} a$ and $a_{y2} = 0.65~\frac{\sqrt{3}}{2} a$, and reduces the reflectance of the structure from  $R = 0.945$ to $R = 1.96\times 10^{-4}$. 
The results in Fig.~\ref{fig:pracoat} agree well with data calculated by a highly accurate multipole scattering matrix method: the RMS difference is $3.4 \times 10^{-3}$, and the only noticeable differences occur on the two sharp resonant features near the lower edge of the figure. Specifically, the multipole-based calculations show that the coating reduces the PC's reflectance from $R=0.943$ to $R=4.29 \times 10^{-4}$.
 
\subsection{All-polarization antireflection coating} 
\label{sub:park_coating}
Finally, we apply our methods to find an all-polarization antireflection coating for a silicon-based self-collimating square-lattice photonic crystal presented by Park \emph{et al.} \cite{Park:2010p651}.  They investigated this class of structures using a scalar treatment of reflections, and were able to design an all-polarization coating at $a/\lambda=0.28$, below the first Wood anomaly.  Since their scalar treatment does not support multiple propagating or evanescent Bloch modes, it generally does not work above the Wood anomaly.  Our method does not have this limitation and we demonstrate this by designing an antireflection coating for both polarizations at a frequency well above the Wood anomaly, using more than one Bloch mode.

Park \emph{et al.} \cite{Park:2010p651} showed that at $a/\lambda = 0.368$, a 2D silicon ($n=3.518$) PC with $r = 0.45~a$ is self-collimating for both polarizations at normal incidence.  The large radius is an extreme case that is challenging to simulate accurately.  At this frequency $M_{\text{min}} = 3$, so for $E_z$ polarized light we include $M = 3$ modes in our calculations, with light incident from the $p=1$ grating order so that the otherwise uncoupled mode is excited.  For $H_z$ light, this procedure does not yield accurate results---Bloch factors are calculated accurately, but the calculated reflection coefficients differ from those calculated directly in COMSOL.  The calculated impedances prove sufficiently accurate to design an effective antireflection coating, but the inaccuracies mean that the coating is not optimal.

To avoid these inaccuracies in $H_z$ polarization, we exploit the symmetry that causes the uncoupled mode.  The physical structure and normally incident field are both symmetric about the $y$-axis, and so modes without even symmetry are not coupled to.  Therefore we formally ignore the uncoupled odd mode, in each PC and in the reference medium, setting $M = 2$.  In our $H_z$ COMSOL simulations for this structure, light is normally incident.

In Fig.~2 of Park \emph{et al.}'s paper \cite{Park:2010p651}, they state that $R\simeq 0.28$ for $E_z$ polarized light, and $R\simeq 0.35$ for $H_z$ light.  We calculate with BlochCode that a semi-infinite slab of the PC has $R= 0.284$ for $E_z$, and $R= 0.354$ for $H_z$ polarized light at this frequency, when incident from silicon.  Specialized FEM-based transfer-matrix calculations agree, showing $R = 0.284$ for $E_z$ polarization, and $R = 0.357$ for $H_z$ polarization.

At $a/\lambda = 0.368$, normally incident light is reflected by the PC into three propagating diffraction orders.  Due to the symmetries of the problem, the $\pm1$ orders are only excited in an even superposition, so light is reflected into two modes.  A successful coating needs to suppress reflection into both these modes simultaneously, and so must balance two modes' amplitudes and two modes' phases simultaneously for each polarization.  Thus the design of a perfect all-polarization coating requires 8 continuous degrees of freedom.  Rather than trying to search an 8-dimensional parameter space, which is computationally expensive even when the evaluation of each point is efficient, we consider coatings with four degrees of freedom and accept that we are unlikely to find an all-polarization coating with zero reflectance.  Nevertheless, this is a particularly difficult problem: not only do we need many degrees of freedom to find a satisfactory coating, but if either of the Bloch factors in a PC is incorrect or any element of the PC's impedance matrix is wrong, then the calculated net reflection off the structure is incorrect as well.

\begin{figure}[htbp]
	\centering
	\includegraphics[]{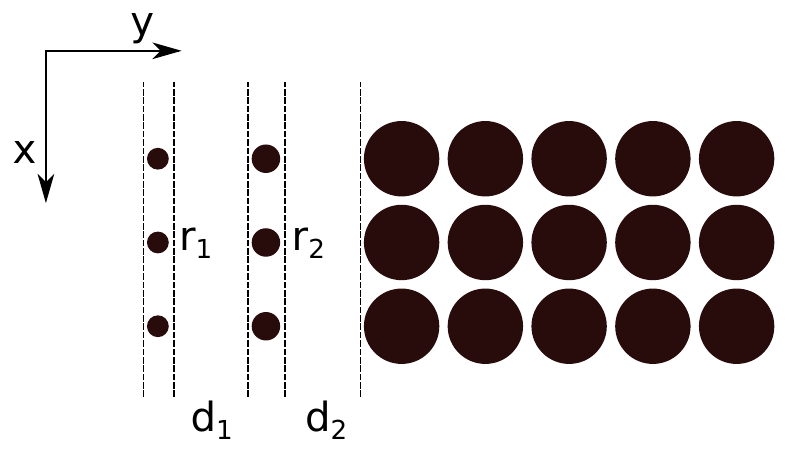}
	\caption{Schematic of the all-polarization antireflection coating.  $r_1$ and $r_2$ are the radii of the holes in the first two layers, and $d_1$ and $d_2$ are the thicknesses of the extra silicon background layers between the first few rows of holes.  For this coating, $r_1 = 0.13~a$, $d_1 = 0.89~a$, $r_2 = 0.17~a$, and $d_2 = 0.9~a$.}
	\label{fig:dualpolschem}
\end{figure}

To limit the coating's thickness, we embed the four degrees of freedom into two rows of holes by varying both the hole radii, $r_1$ and $r_2$, and the space after the layers, $d_1$ and $d_2$ (Fig.~\ref{fig:dualpolschem}).  Increasing $d_1$ and $d_2$ is similar to increasing $a_y$, as in Sec.~\ref{sub:oldcoating}, but because the candidate PCs are independent of $d$, only one PC per radius needs to be simulated in COMSOL.  Furthermore, the properties of the layers of silicon with thickness $d_i$ may be calculated analytically.  We consider 36 possible hole radii in the range $r_i \in [0.10,0.45]~a$ and 99 values of $d_i \in (0,1)~a$.  To allow a thin coating, we set $a_y = 2r + 0.1a$ for each PC.  If necessary, additional degrees of freedom could be added to find a coating with even lower reflectances.

On a single core of a $16 \times 2.4$~GHz Intel Xeon-Quad workstation, it took a total of 15 minutes to find the modes of the 36 PCs in the two polarizations.  For $E_z$ polarization, $w^2 \simeq 10^{-5}$ for most radii, and for $H_z$ polarization $w^2$ ranged roughly from $3 \times 10^{-3}$ for thin unit cells to $10^{-7}$ for the thicker cells with larger radius. Due to the large number of candidate coatings ($\sim 1.3 \times 10^7$), the embarrassingly parallel problem was split over 16 cores of the workstation, taking approximately 80 minutes per polarization.

The best $E_z$ coating reduces $R$ from 0.284 to $9.56 \times 10^{-5}$, and the best $H_z$ coating reduces $R$ from 0.354 to $3.33 \times 10^{-4}$.  The best all-round coating is taken to be the one with the lowest total reflection in the two polarizations.  This coating has $r_1 = 0.13~a$, $d_1 = 0.89~a$, $r_2 = 0.17~a$, and $d_2 = 0.90~a$ (Fig.~\ref{fig:dualpolschem}).  In $E_z$ it reduces $R$ to 0.0141, and in $H_z$ it reduces $R$ to 0.0197.  Calculations from a specialized transfer matrix method \cite{Botten:2004p5} agree with these results, giving $R = 0.0142$ in $E_z$ polarization and $R = 0.0211$ in $H_z$.  

To verify these results without the aid of our specialized methods, implementations of which are not publicly available, we simulate the structure using COMSOL Multiphysics.  Since COMSOL cannot directly calculate reflection coefficients off semi-infinite PCs, we simulate a 20-period section of the uncoated PC surrounded by the background dielectric, and compare the results to a simulation with the antireflection coating on both sides of the PC section.  BlochCode calculates the reflectance of the uncoated and coated structures to be 0.407 and 0.0124 respectively in the $E_z$ polarization, and 0.574 and 0.0074 in the $H_z$ polarization.  The COMSOL simulations agree with these results, showing that the coating reduces $R$ from 0.407 to 0.0129 in the $E_z$ polarization, and from 0.585 to 0.0055 in the $H_z$ polarization.
\section{Discussion \& Conclusion} 
\label{sec:conclusion}
We have detailed a method for calculating the complex band structure and impedance of PCs.  The method takes into account structural symmetries in the PC, and enforces relationships between the fields of forward and backward modes, thus improving the method's accuracy by eliminating ill-conditioning and constraining modal fields.  We have applied the method to three cases, and have demonstrated that it works for a variety of square and triangular lattice 2D photonic crystals, for light in both polarizations and at different incident angles.  We have demonstrated that our method works at frequencies both above and below the first Wood anomaly, the frequency above which scalar methods cannot adequately describe light propagation and reflection in PCs.

The stronger the excitation of a Bloch mode, the more accurately our method calculates its properties.  Thus the method is well-suited to calculating reflection and transmission through arbitrary PC stacks, where the most important modes are those that are strongly excited.  Since PC impedances make it so easy to calculate the reflection and transmission properties of many combinations of PCs in a stack, it is feasible to search large parameter spaces of PC stacks for particular reflective properties over a range of frequencies, incident angles and polarizations.  The method can be used to design not only all-polarization antireflection coatings, but also broadband antireflection coatings \cite{Lawrence:2009p11}, polarization filters, angular filters, and other devices.

Ha \emph{et al.} have applied their method to slab PC waveguides \cite{Ha:2011p2082}. 
We have not yet applied our method to any 3D structure. 
As long as the $x-z$ plane mirror symmetry is present, our method for finding the complex band structure remains valid. 
The field of a slab waveguide might be sampled only over the PC's surface (as in a SNOM experiment \cite{Ha:2011p2082}) or throughout the entire volume of the structure (as in a simulation); either case provides sufficient information to determine the modal fields within the sampled region and the associated complex band structure. 
However, the impedance formalism is yet to be developed for 3D structures.

Our method is also valid for finding modes of PC waveguides, using supercells. Calculation of reflection and transmission matrices between PC waveguides is yet to be demonstrated using impedances, but they have previously been calculated directly from the supercell's $\mathbf{E}$ and $\mathbf{H}$ matrices \cite{deSterke:09}.

Bloch mode analysis is a valuable tool in understanding light's interactions with PCs.  Using an EM solver and our method, for which source code is available \cite{blochcodeurl}, it is straightforward to find a PC's complex band structure and its impedance.  Respectively, these quantities dictate how the Bloch modes travel through the PC, and which modes they couple with at a PC interface.  If these quantities are known for a set of PCs, then it is fast and efficient to calculate how light travels through arbitrary stacks of the PCs.

\acknowledgments
This research was conducted by the Australian Research Council Centre of Excellence for Ultrahigh bandwidth Devices for Optical Systems (project number CE110001018).

\end{document}